# Carbonaceous dust grains seen in the first billion years of cosmic time

Joris Witstok[1,2], Irene Shivaei[3*], Renske Smit[4*], Roberto Maiolino[1,2,5], Stefano Carniani[6], Emma Curtis-Lake[7], Pierre Ferruit[8], Santiago Arribas[9], Andrew J. Bunker[10], Alex J. Cameron[10], Stephane Charlot[11], Jacopo Chevallard[10], Mirko Curti[1,2,12], Anna de Graaff[13], Francesco D'Eugenio[1,2], Giovanna Giardino[14], Tobias J. Looser[1,2], Tim Rawle[15], Bruno Rodríguez del Pino[9], Chris Willott[16], Stacey Alberts[3], William M. Baker[1,2], Kristan Boyett[17,18], Eiichi Egami[3], Daniel J. Eisenstein[19], Ryan Endsley[20], Kevin N. Hainline[3], Zhiyuan Ji[3], Benjamin D. Johnson[19], Nimisha Kumari[21], Jianwei Lyu[3], Erica Nelson[22], Michele Perna[9], Marcia Rieke[3], Brant E. Robertson[23], Lester Sandles[1,2], Aayush Saxena[5,10], Jan Scholtz[1,2], Fengwu Sun[3], Sandro Tacchella[1,2], Christina C. Williams[24] & Christopher N. A. Willmer[3]

**Large dust reservoirs (up to ~$10^8$ M$_\odot$) have been detected[1-3] in galaxies out to redshift z ~ 8, when the age of the Universe was only about 600 Myr. Generating significant amounts of dust within such a short timescale has proven challenging for theories of dust formation[4,5] and has prompted the revision of the modelling of potential sites of dust production[6–8] such as the atmospheres of asymptotic giant branch (AGB) stars in low-metallicity environments, supernovae (SNe) ejecta, and the accelerated growth of grains in the interstellar medium (ISM). However, degeneracies between different evolutionary pathways remain when the total dust mass of galaxies is the only available observable. Here we report observations of the 2175 Å dust attenuation feature, well known in the Milky Way (MW) and galaxies at $z \lesssim 3$[9–11], in the near-infrared spectra of galaxies up to $z$ ~ 7, corresponding to the first billion years of cosmic time. The relatively short timescale implied for the formation of carbonaceous grains giving rise to this feature[12] suggests a rapid production process, likely in Wolf-Rayet (WR) stars or SN ejecta.**

As part of the *JWST* Advanced Deep Extragalactic Survey (JADES), we obtained deep Near-Infrared Spectrograph (NIRSpec) multi-object spectroscopy taken in the PRISM configuration (spectral range 0.6 μm to 5.3 μm, resolving power $R$ ~ 100). Using the NIRSpec micro-shutter array (MSA), we observed 253 sources across three visits between 21 and 25 October 2022 (*JWST* programme 1210; PI: Lützgendorf), with exposure times per object ranging from 9.3 to 28 hours. The extracted one-dimensional spectra reached a continuum sensitivity ($3\sigma$) of ~6-40 × $10^{-22}$ erg $s^{-1}$ $cm^{-2}$ $Å^{-1}$ (~27.2-29.1 AB magnitude) at ~2 μm. Targets were selected with a specific focus on high-redshift galaxies in imaging with the *Hubble Space Telescope* (*HST*) and *JWST*/Near-Infrared Camera (NIRCam).

[1]Kavli Institute for Cosmology, University of Cambridge, Madingley Road, Cambridge, CB3 0HA, UK. [2]Cavendish Laboratory, University of Cambridge, 19 JJ Thomson Avenue, Cambridge, CB3 0HE, UK. [3]Steward Observatory, University of Arizona, 933 N. Cherry Avenue, Tucson AZ 85721, USA. [4]Astrophysics Research Institute, Liverpool John Moores University, 146 Brownlow Hill, Liverpool L3 5RF, UK. [5]Department of Physics and Astronomy, University College London, Gower Street, London WC1E 6BT, UK. [6]Scuola Normale Superiore, Piazza dei Cavalieri 7, I-56126 Pisa, Italy. [7]Centre for Astrophysics Research, Department of Physics, Astronomy and Mathematics, University of Hertfordshire, Hatfield AL10 9AB, UK. [8]European Space Agency, European Space Astronomy Centre, Madrid, Spain. [9]Centro de Astrobiología (CAB), CSIC–INTA, Cra. de Ajalvir Km. 4, 28850- Torrejón de Ardoz, Madrid, Spain. [10]University of Oxford, Denys Wilkinson Building, Keble Road, Oxford OX1 3RH, UK. [11]Sorbonne Université, CNRS, UMR 7095, Institut d'Astrophysique de Paris, 98 bis bd Arago, 75014 Paris, France. [12]European Southern Observatory, Karl-Schwarzschild-Strasse 2, D-85748 Garching bei Muenchen, Germany. [13]Max-Planck-Institut für Astronomie, Königstuhl 17, D-69117, Heidelberg, Germany. [14]ATG Europe for the European Space Agency, ESTEC, Noordwijk, The Netherlands. [15]European Space Agency, Space Telescope Science Institute, Baltimore, Maryland, USA. [16]NRC Herzberg, 5071 West Saanich Rd, Victoria, BC V9E 2E7, Canada. [17]School of Physics, University of Melbourne, Parkville 3010, VIC, Australia. [18]ARC Centre of Excellence for All Sky Astrophysics in 3 Dimensions (ASTRO 3D), Australia. [19]Center for Astrophysics | Harvard & Smithsonian, 60 Garden St., Cambridge MA 02138 USA. [20]Department of Astronomy, University of Texas, Austin, TX 78712, USA. [21]AURA for European Space Agency, Space Telescope Science Institute, 3700 San Martin Drive, Baltimore MD 21210, USA. [22]Department for Astrophysical and Planetary Science, University of Colorado, Boulder, CO 80309, USA. [23]Department of Astronomy and Astrophysics University of California, Santa Cruz, 1156 High Street, Santa Cruz CA 96054, USA. [24]NSF's National Optical-Infrared Astronomy Research Laboratory, 950 North Cherry Avenue, Tucson, AZ 85719, USA.
*These authors contributed equally to this work.
Corresponding authors: Joris Witstok (jnw30@cam.ac.uk), Irene Shivaei (ishivaei@arizona.edu), Renske Smit (r.smit@ljmu.ac.uk).

Through visual inspection of all spectra, we find strong evidence of the absorption feature around a rest-frame wavelength $\lambda_{emit} = 2175$ Å in the spectrum of a galaxy at $z = 6.71$ (JADES-GS+53.15138-27.81917; JADES-GS-z6-0 hereafter), revealed via a significant ($6\sigma$) deviation from a smooth power-law continuum, as shown in Fig. 1. This feature, known as the UV attenuation "bump", was first discovered by Stecher (1965) along sightlines in the MW[9] and is attributed to carbonaceous dust grains, specifically polycyclic aromatic hydrocarbons (PAHs) or nano-sized graphitic grains[12]. We fitted a Drude profile around 2175 Å to the excess attenuation[13], defined as the observed spectrum normalised to a bump-free attenuated spectrum that is predicted by a power-law function fitted outside of the bump region (see Methods). We find a bump strength (amplitude) of $0.43^{+0.07}_{-0.07}$ mag and a central wavelength $\lambda_{max} = 2263^{+20}_{-24}$ Å. The latter, while within the range expected by models of carbonaceous grains[14], is notably higher than the range typically observed along different sightlines in the MW, potentially suggestive of a change in grain mixture[15]. Beyond the local Universe, the feature has previously only been observed spectroscopically in massive, metal-enriched galaxies at $z \lesssim 3$, suggesting it originates in dust grains exclusively present in evolved galaxies[11,13,16–19]. The detection reported here is the first direct, spectroscopic detection of the UV bump in galaxies at $z > 3$.

The properties of JADES-GS-z6-0 are summarised in Table 1. This galaxy shows significant dust obscuration as probed by the ratio of Balmer lines ("Balmer decrement"): here, Hα/Hβ ~ 3.7 indicates a nebular extinction of $E(B − V)_{neb} = 0.25 \pm 0.07$ mag. In agreement with the trend between metallicity and bump strength observed at lower redshift, measurements of the gas-phase and stellar metallicity ($Z$ ~ 0.2-0.3 $Z_\odot$) further suggest JADES-GS-z6-0 has undergone substantial metal enrichment relative to galaxies with similar mass at the same redshift[20].

To systematically investigate the prevalence of the UV bump and obtain clues on its origin at such early times, we further selected JADES galaxies on the basis of a confident spectroscopic redshift above $z > 4$ with a median signal-to-noise ratio (SNR) of at least 3 per spectral pixel in the region corresponding to rest-frame wavelengths of 1268 Å $< \lambda_{emit} <$ 2580 Å. This results in a sample of 49 objects between redshift 4.02 and 11.48. Comparing the continuum slopes on both sides of the central wavelength at 2175 Å (see Methods), we selected ten galaxies (at $4.02 < z < 7.20$) from this parent sample whose spectral shape points towards the presence of a UV bump.

We constructed a weighted average ("stack"; see Methods) of all 49 objects in our parent sample as well as our ten selected objects with evidence for a bump signature, as shown in Fig. 2. In both stacks, we find evidence for emission from the C III $\lambda$ 1907, 1909 Å nebular lines that are commonly seen in metal-poor galaxies[21]. There is no indication of the bump in the parent sample; the stacked spectrum of the ten selected objects, however, shows a clear depression ($5\sigma$) centred on ~2175 Å. Though we do not find evidence for significant differences in stellar properties (i.e. mass or age), these ten galaxies are characterised by a considerable amount of dust obscuration, comparable to $z$ ~ 2 galaxies with the bump feature[18], and mildly enhanced metallicities compared to the parent sample (see Methods). We again fitted a Drude profile to the excess attenuation in the stacked spectrum of these ten objects, finding a bump amplitude of $0.10^{+0.01}_{-0.01}$ mag and a central wavelength $\lambda_{max} = 2236^{+21}_{-20}$ Å.

While the UV bump has long been known to exist, its variable presence and strength has been an open topic of debate in galaxy evolution studies[13,22]. The feature is commonly attributed to PAHs[12], molecules thought to be susceptible to destruction by hard ionising radiation, and it is present in the MW and Large Magellanic Cloud (LMC) extinction curves, but very weak or absent in the Small Magellanic Cloud (SMC) curves[23]. In the attenuation curve of individual galaxies, radiative-transfer effects determined by the dust-star geometry can weaken the bump in the observed integrated spectrum[24,25]. However, by stacking the photometry of large samples of galaxies, the bump has been detected to varying degrees at redshifts $z \lesssim 3$, with tentative hints at $z \lesssim 6$[11,17,19,26]. Spectroscopically, the bump has only been seen in relatively massive and dusty individual galaxies at $z \sim 2$[13,18]. In Fig. 3, the bump amplitude is shown as a function of cosmic time, including its strength in the extinction curves in MW, LMC, and SMC sightlines[23]. Our inferred bump amplitude and central wavelength, especially in the individual spectrum of JADES-GS-z6-0, are comparatively high, the former defying the trend with stellar mass seen at lower redshift. This may point towards a different nature of the grains responsible for the absorption (e.g. graphite instead of PAHs) in addition to a different, likely simpler, dust-star geometry compared to lower-redshift counterparts – intriguingly, there is tentative evidence for a colour gradient in JADES-GS-z6-0 (see Methods).

Moreover, a direct detection of the bump at $z \sim 4$-$7$ is striking given that at these redshifts, the age of the Universe is only around a billion years (~800 Myr at $z = 6.71$). Substantial production of carbon and subsequent formation of carbonaceous grains responsible for the absorption feature through the standard AGB channel, particularly in the low-metallicity regime characterising such early galaxies (i.e. $Z \sim 0.1$ $Z_{\odot}$[20]), would require low-mass ($M \lesssim 2.5$ $M_{\odot}$) and hence long-lived stars to reach the AGB at the end of their lives, after more than 300 Myr[7]. If this is the dominant channel via which carbonaceous grains are formed, the presence of the UV bump implies the onset of star formation in these galaxies occurred within the first half billion years of cosmic time, corresponding to redshift $z \gtrsim 10$. Indeed, star formation has been shown to occur at this early epoch with the confirmation of $z > 10$ galaxies[27]. However, in our sample we do not find evidence for substantial star formation activity that occurred on timescales beyond 300 Myr (see Methods). The absence of clear signatures from such relatively old stellar populations suggests that other, faster channels for the production of carbonaceous dust are required in these early systems, corroborated by the high observed frequency of extremely metal-poor MW stars that are carbon enhanced[28].

One explanation is that these grains are formed on significantly shorter timescales via more massive and rapidly evolving stars, possibly by SNe or WR stars, which would overhaul some, and place strong constraints on other, theoretical models of dust production and stellar evolution. PAH production has indeed been observed in WRs[29], and while subsequent SN type-Ib/c explosions are generally expected to destroy most dust produced in the preceding WR phase, models have shown that carbonaceous grains produced by binary carbon-rich WRs (WCs) can survive[30]. However, for standard initial mass functions (IMFs), WR and in particular WC stars are rare[31]. Conversely, isotopic signatures in presolar graphite grains found on primitive meteorites indicate a type-II SN origin, suggesting the production of these potential carriers of the UV bump starts at early times[32]. Indeed, dust production in SN ejecta has been regarded as a potential rapid channel for significant dust production in the early Universe[33], its net efficiency depending on the grain destruction rate in the subsequent reverse shock[34]. However, substantial *carbonaceous* production in SN ejecta is expected only by some classes of models and for a certain subclass of scenarios (e.g. non-rotating progenitors), while other models favour the formation of silicates or other types of dust[35–38]. In summary, our detection of carbonaceous dust at $z \sim 4\text{-}7$ provides crucial constraints on the dust production models and scenarios in the early Universe.

**Table 1 | Properties of JADES-GS-z6-0.**

| | |
|---|---|
| RA (deg) | +53.15138 |
| Dec (deg) | −27.81917 |
| $t_{\rm exp}$ (h) | 27.9 |
| $z_{\rm spec}$ | $6.70647^{+0.00044}_{-0.00033}$ |
| $m_{\rm F115W}$ (mag) | $28.58 \pm 0.12$ |
| $M_{\rm UV}$ (mag) | $-18.34 \pm 0.12$ |
| $\beta_{\rm UV}$ | $-2.13^{+0.18}_{-0.18}$ |
| $\gamma_{34}$ | $-3.6^{+1.5}_{-1.2}$ |
| $Z_{\rm neb}$ ($Z_\odot$) | $0.17^{+0.05}_{-0.04}$ |
| $E(B-V)_{\rm neb}$ (mag) | $0.25 \pm 0.07$ |
| $M_*$ ($10^8\ M_\odot$) | $1.0^{+0.3}_{-0.2}$ |
| $Z_*$ ($Z_\odot$) | $0.34^{+0.05}_{-0.05}$ |
| SFR$_{30}$ ($M_\odot$ yr$^{-1}$) | $3.0^{+2.0}_{-1.0}$ |
| $t_*$ (Myr) | $18^{+11}_{-7}$ |

Error bars represent a $1\sigma$ uncertainty. Rows: (1) Right Ascension (RA) in J2000, (2) Declination (Dec) in J2000, (3) Exposure time ($t_{\rm exp}$) in the NIRSpec PRISM spectra in hours, (4) Spectroscopic redshift ($z_{\rm spec}$), (5) Apparent AB magnitude in the NIRCam F115W filter ($m_{\rm F115W}$), (6) Absolute AB magnitude in the UV ($M_{\rm UV}$), (7) UV spectral slope ($\beta_{\rm UV}$), (8) Spectral slope change around $\lambda_{\rm emit} = 2175$ Å ($\gamma_{34}$), (9) Gas-phase metallicity ($Z_{\rm neb}$; from rest-frame optical emission lines) in units of Solar metallicity, (10) Nebular extinction ($E(B-V)_{\rm neb}$; from the Balmer decrement) in magnitudes, (11) Stellar mass ($M_*$) in $10^8$ Solar masses, (12) Stellar metallicity ($Z_*$; from SED modelling) in units of Solar metallicity, (13) Star formation rate in Solar masses per year averaged on a timescale of 30 Myr (SFR$_{30}$), (14) Mass-weighted stellar age ($t_*$) in Myr.

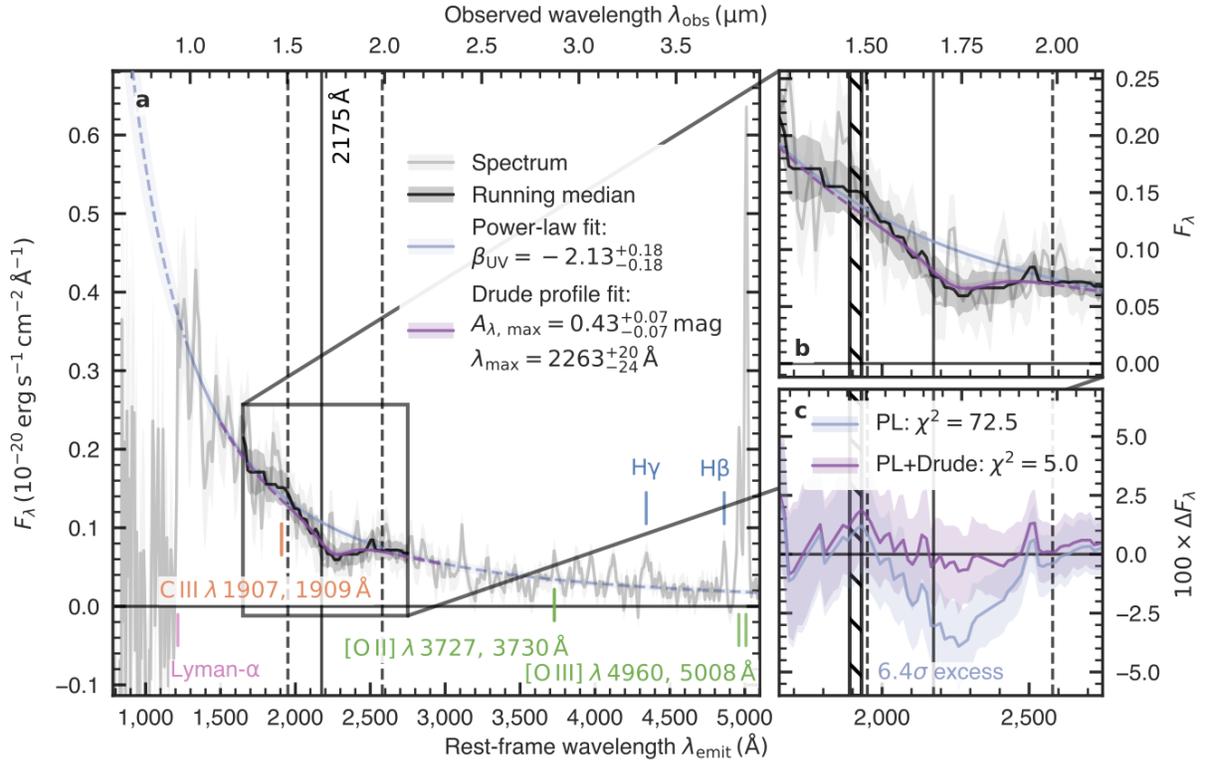

**Fig. 1 | Spectrum taken by *JWST*/NIRSpec of JADES-GS-z6-0 at redshift $z = 6.71$. a**, Overview of the spectrum (grey solid line) with a power-law fit to the UV continuum (blue solid line). Several spectral features used for spectroscopic redshift confirmation are indicated, including the Lyman-α break, the [O II] $\lambda\, 3727, 3730$ Å doublet, the Hβ, Hγ and [O III] $\lambda\, 4960, 5008$ Å lines. **b**, Zoom-in of the UV bump region around $\lambda_{emit} = 2175$ Å where a running median (solid black line), representing the attenuated stellar continuum, reveals a deep localised absorption profile. A Drude profile fit within the vertical dashed lines (purple solid line) with respect to the smooth power law (blue solid line) yields an amplitude of $0.43^{+0.07}_{-0.07}$ mag and a central wavelength $\lambda_{max} = 2263^{+20}_{-24}$ Å. The hatched region indicates the C III $\lambda\, 1907, 1909$ Å doublet. **c**, The residuals ($\Delta F_\lambda$) show that the power-law (PL) fit alone has a significant negative flux excess between ~ 2000 Å and 2400 Å ($6.4\sigma$), while the power-law and Drude profile combined (PL+Drude; purple line) provides a significantly better fit (chi squared of $\chi^2 = 72.5$ versus $\chi^2 = 5.0$ for PL and PL+Drude, respectively). All shading represents $1\sigma$ uncertainty.

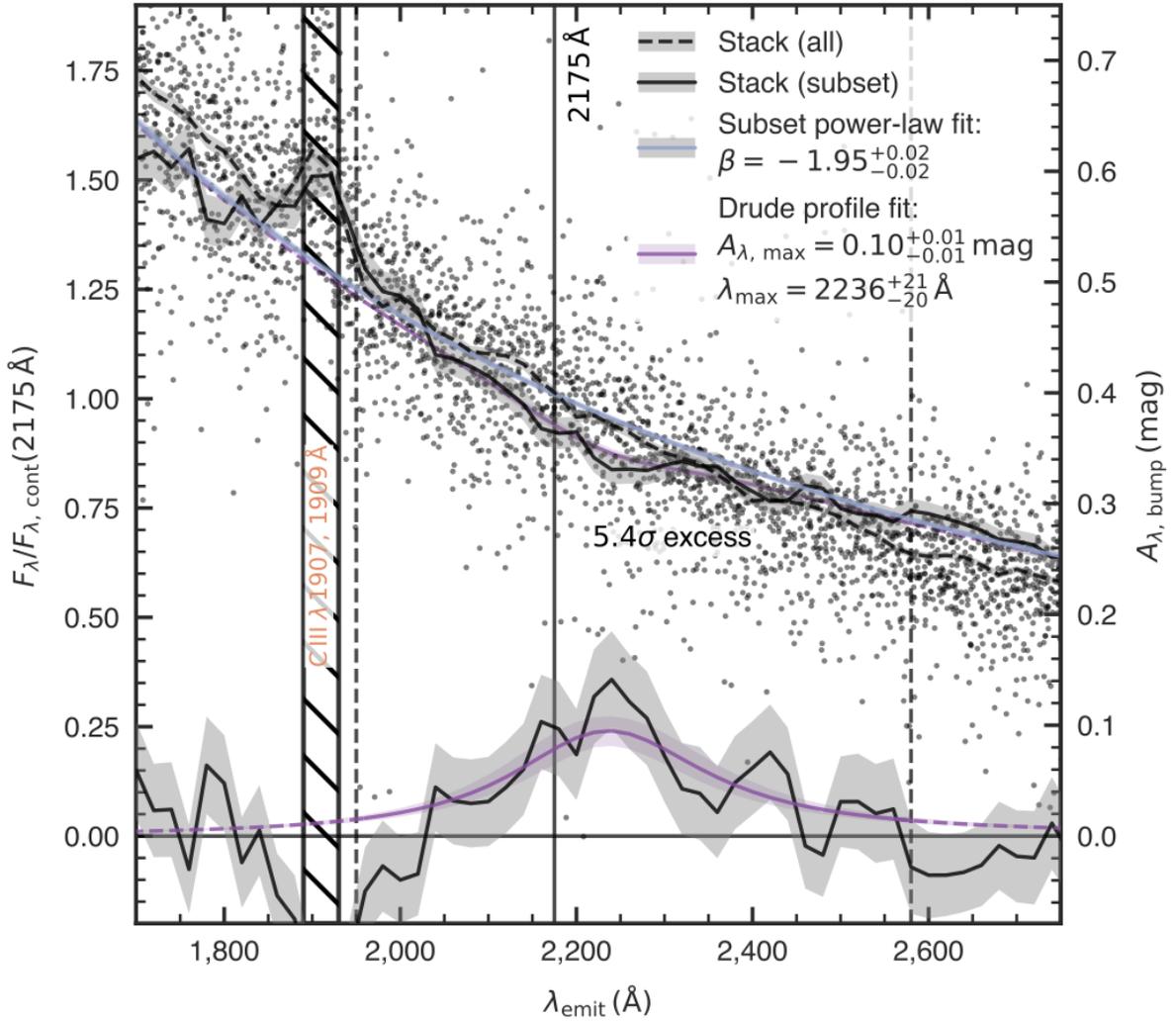

**Fig. 2 | Normalised and stacked spectra around the UV bump of z > 4 JADES galaxies observed by *JWST*/NIRSpec.** Spectra of all galaxies (small black dots) are shifted to the rest frame and normalised to the predicted continuum level at a rest-frame wavelength of $\lambda_{emit} = 2175$ Å in the absence of a UV bump (see Methods section). The dashed black line (shading represents 1σ uncertainty) shows a stacked spectrum obtained by combining all 49 objects in wavelength bins of $\Delta\lambda_{emit} = 20$ Å, clearly revealing emission from the C III $\lambda$ 1907, 1909 Å doublet in the hatched region. The stacked spectrum of ten galaxies selected to have a bump signature (solid black line, shading as 1σ uncertainty), in addition to appearing to have a mildly redder UV slope, shows the presence of the UV bump around 2175 Å having an excess with respect to a power-law continuum (solid blue line; see Methods) at a significance of 5.4σ. The excess attenuation $A_{\lambda,\,bump}$ (curve at the bottom, corresponding to the axis on the right) is fitted with a Drude profile (shown in purple with shading as 1σ uncertainty), where we find an amplitude of $0.10^{+0.01}_{-0.01}$ mag and a central wavelength $\lambda_{max} = 2236^{+21}_{-20}$ Å.

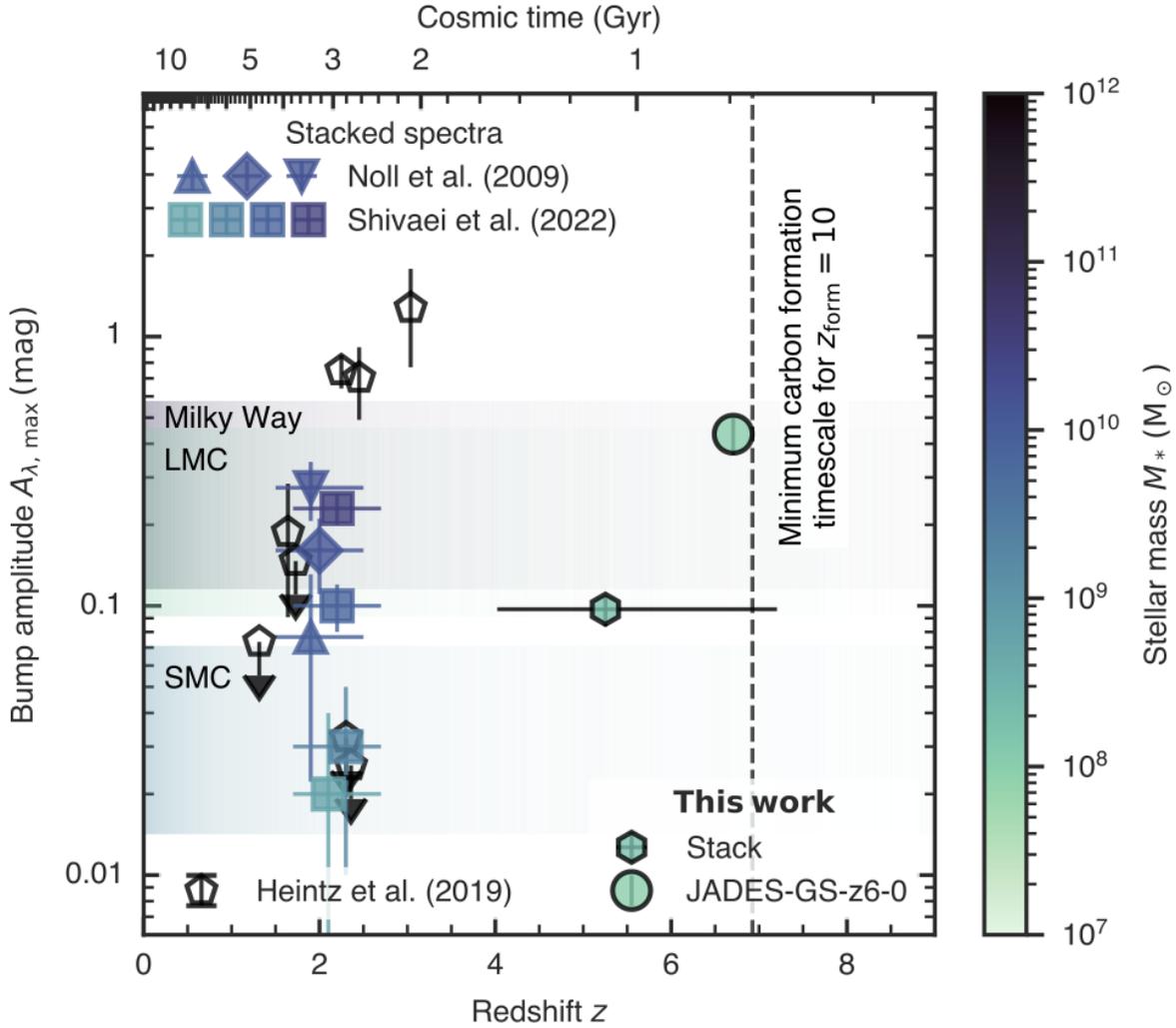

**Fig. 3 | Redshift evolution of UV bump strength.** The amplitude of the excess attenuation, $A_{\lambda,\,max}$, is shown for JADES-GS-z6-0 individually as well as for the stack of ten $z \sim 4\text{-}7$ JADES galaxies. Points are coloured according to their (average) stellar mass; error bars along the y-axis represent $1\sigma$ uncertainty. At $z \sim 2$, measurements from gamma-ray burst absorbers (Heintz et al. and references therein) and from stacked spectra in various bins of stellar mass (Shivaei et al.) or shape of the UV continuum as a whole and in the bump region (Noll et al.) are shown (see Methods for details)[13,18,39]. Error bars of the stacked spectra along the x-axis represent the full redshift range, their central values slightly shifted for visualisation purposes. The bump amplitudes in the average MW, LMC, and SMC dust extinction curves[23,40], converted to an attenuation for a visual extinction range of 0.1 mag $< A_V <$ 0.5 mag, are indicated with light shadings. The age of the Universe is indicated at the top. A vertical dashed line indicates the minimum timescale required for carbon production by AGB stars (i.e. 300 Myr) if the galaxy formed at $z_{form} = 10$.

# Methods

## Data and parent sample

The observations presented here were taken as part of JADES[41], a joint survey conducted by the *JWST*[42,43] NIRCam[44] and NIRSpec[45,46] Guaranteed Time Observations (GTO) instrument science teams. As described in Robertson et al.[47] and Curtis-Lake et al.[27], deep NIRCam imaging[48] over a wavelength range $\lambda_{obs} \simeq 0.8$ μm to 5 μm (reaching $m_{AB}$ ~ 30 mag in F200W) was taken under *JWST* programme 1180 (PI: Eisenstein) in an area of 65 arcmin$^2$ over the Great Observatories Origins Deep Survey – South (GOODS-S), which includes the Hubble Ultra Deep Field (HUDF). We additionally make use of public medium-band imaging taken as part of the *JWST* Extragalactic Medium-band Survey (JEMS[49]; *JWST* programme 1963, PI: Williams) and First Reionization Epoch Spectroscopic COmplete Survey (FRESCO[50]; *JWST* programme 1895, PI: Oesch). Incorporating the wealth of publicly available ancillary data from *HST*, a catalogue with photometric redshifts was constructed to identify high-redshift galaxy candidates. NIRSpec multi-object spectroscopy[51] of these NIRCam-selected sources was taken using the MSA[52] in the PRISM/CLEAR spectral configuration, covering a spectral range 0.6 μm to 5.3 μm with resolving power $R$ ~ 100. A three-point nodding pattern was implemented for background subtraction in addition to small dithers with MSA reconfigurations to increase sensitivity and flux accuracy, improve spatial sampling, mitigate the impact of the detector gaps, and aid the removal of cosmic rays. Dither pointings consisted of four sequences of three nodded exposures. Each setup was made up of two integrations of 19 groups, resulting in exposure times of 8403.2 seconds for each sequence and of 33612 seconds (9.3 hours) for each dither pointing[27]. The main galaxy considered in this work (JADES-GS-z6-0) was observed in three visits, resulting in an integration time of 27.9 hours (Table 1), while other targets had exposure times ranging between 9.3 hours and 27.9 hours. The spectral energy distribution (SED) and a false-colour image of JADES-GS-z6-0 with the location of the NIRSpec MSA shutters overlaid is shown in Extended Data Fig. 1. We note the imaging reveals a tentative colour gradient, with the shutter capturing the redder part of the galaxy which may contribute to the strength of the UV bump in the spectrum of JADES-GS-z6-0.

Flux-calibrated two-dimensional spectra and one-dimensional spectral extractions were obtained with pipelines developed by the ESA NIRSpec Science Operations Team (SOT) and the NIRSpec GTO team, which will be discussed in detail in a forthcoming paper. The pipeline generally adopts the same algorithms included in the official STScI pipeline that generates MAST archive products. An irregular wavelength grid with 5 spectral pixels per resolution element was adopted to avoid oversampling of the line spread function at short wavelengths ($\lambda_{obs}$ ~ 1 μm). Extraction of the one-dimensional spectra was performed with a 5-pixel aperture covering the entire shutter size to recover all emission. However, as in Curtis-Lake et al.[27], we considered an additional extraction over a 3-pixel aperture to test the robustness of our findings, as will be discussed in the Robustness of the UV bump detections section. Given the compact sizes of the high-redshift galaxies considered here (see Extended Data Fig. 1), slit-loss corrections were applied under the assumption of a point-like source placed at the relative intra-shutter position of each galaxy. We note that systematic uncertainties in the slit-loss correction will be a smooth function of wavelength and will therefore not affect the UV bump signature, which instead relies on the detection of the UV slope inflection over a relatively small wavelength range around $\lambda_{emit} = 2175$ Å (as discussed in the next sections). Extraction was performed in a shutter-size aperture to recover all emission. Further details regarding the target selection and data reduction are extensively discussed in preceding JADES works[27,41,47,48,51].

**Sample selection**

After an automated spectral fitting routine with BAGPIPES (Bayesian Analysis of Galaxies for Physical Inference and Parameter EStimation) code[53], spectroscopic redshift estimates were confirmed by visual inspection independently by at least two team members. The final redshift values were determined by a subsequent analysis (described in detail in Chevallard et al., in prep.) with the BEAGLE (BayEsian Analysis of GaLaxy sEds) code[54], as described in Curtis-Lake et al.[27] but with a star-formation history (SFH) consisting of a 10 Myr long star-formation burst combined with a delayed exponential component, and a narrow redshift prior distribution centred around the visually confirmed redshifts. We selected objects on the basis of a confident spectroscopic redshift above $z > 4$ to ensure the rest-frame UV coverage includes the Lyman-α break. Based on the formal uncertainty, we further selected spectra with median SNR of at least 3 in the region corresponding to rest-frame wavelengths of 1268 Å < $\lambda_{emit}$ < 2580 Å.

We then performed several Bayesian power-law fitting procedures to the rest-frame UV continuum with a Python implementation[55] of the MultiNest[56] nested sampling algorithm. To identify spectra exhibiting a UV bump, we fitted power laws in four adjacent wavelength windows defined by Noll et al.[57] (with corresponding power-law indices $\gamma_1$ to $\gamma_4$), excluding the region 1920 Å < $\lambda_{emit}$ < 1950 Å to avoid contamination by the C III doublet. In the presence of the UV bump, the spectral shape of the rest-frame UV is characterised by a strong turnover in power-law slope directly blue- and redwards of 2175 Å covered by regions 3 and 4 respectively, resulting in a negative $\gamma_{34} \equiv \gamma_3 - \gamma_4$ value. Before fitting these separate wavelength windows in the individual spectrum, we applied a running median filter over 15 spectral pixels that cover 3× the spectral resolution. We estimated the uncertainty on the running median with a bootstrapping procedure where we randomly perturb each of the 15 spectral pixels according to their formal uncertainty for 100 iterations.

In the fitting algorithm, a likelihood was calculated based on the inverse-variance weighted squared residuals between a given model and the observed spectrum within the adopted spectral regions. We chose flat prior distributions for the power-law indices (ranging between $-5 < \gamma_i < 1$) and normalisation at the centre of each wavelength window (between 0 and twice the maximum value of the spectrum in the fitting regions). Best-fit values of $\gamma_{34}$, whose posterior distribution was obtained from simultaneously fitting $\gamma_3$ and $\gamma_4$, are shown in Extended Data Fig. 2 as the 50th percentile (i.e. median) with 16th and 84th percentiles as a ±1$\sigma$ confidence range. A selection of galaxies with median value of $\gamma_{34} < -1$, in addition to $\gamma_{34} < 0$ within the 1$\sigma$ uncertainty range, led to the identification of ten galaxies (including JADES-GS-z6-0) with evidence for a UV bump (the "bump sample"). The next section will discuss the physical properties of this subsample in the context of the full sample. Coordinates and other properties of these ten galaxies are reported in Extended Data Table 1.

## Physical properties

We consistently used a flat ΛCDM cosmology based on the results of the Planck collaboration[58] (i.e. $H_0 = 67.4$ km s$^{-1}$ Mpc$^{-1}$, $\Omega_m = 0.315$) throughout. Several of the main physical properties of the full sample are presented in Extended Data Fig. 2. Extended Data Table 1 lists observed properties of the ten individual galaxies in the bump sample, while Extended Data Table 2 reports median values for the bump sample, the sample of galaxies not contained in the bump sample (the "non-bump sample"), and the full sample, as well as values measured from the stacked spectra.

## UV magnitudes and slopes

We derived UV magnitudes directly from NIRCam photometric data points probing a rest-frame wavelength of ~1500 Å (F115W in the case of JADES-GS-z6-0; see Table 1), if available (we note several targets fall outside of the NIRCam footprint). We fitted an overall UV slope $\beta_{UV}$ to the rest-frame UV continuum probed by the NIRSpec PRISM measurements using a similar Bayesian power-law fitting procedure as described in the Sample selection section. We adopted the spectral windows defined by Calzetti et al.[59] which are designed to exclude several UV emission and absorption features. Indeed, no strong emission lines are observed within these spectral regions of our low-resolution spectra, and importantly they explicitly exclude the bump region and C III emission lines. We chose a Gaussian prior distribution for the power-law index (centred on $\mu_\beta = -2$ with a width of $\sigma_\beta = 0.5$) and a flat prior on the normalisation at $\lambda_{emit} = 1500$ Å (between 0 and twice the maximum value of the spectrum in the fitting regions). The resulting UV slope of JADES-GS-z6-0 is reported in Table 1.

## Spectroscopic rest-frame optical properties

Emission line fluxes in the NIRSpec PRISM measurements of individual galaxies in our sample were obtained using the pPXF software[60] (for details, we refer to Curti et al.[20]). We converted Hα/Hβ line ratios into a nebular extinction $E(B-V)_{neb}$ under the Cardelli et al.[61] extinction curve, assuming an intrinsic ratio of Hα/Hβ = 2.86 appropriate for case-B recombination, $T_e = 10^4$ K, and $n_e = 100$ cm$^{-3}$ (e.g. ref.[62]). We note that for JADES-GS+53.13423-27.76891 at $z = 7.0493$ the Hα line is located precisely on the edge the PRISM spectral coverage, causing the measured Hα/Hβ ratio to appear significantly below the theoretical value of Hα/Hβ = 2.86 expected in the absence of dust. Moreover, we caution that potential wavelength-dependent slit-loss effects could bias the Hα/Hβ measurement (although minimally as the objects in this analysis are not significantly resolved) and that the stellar and nebular extinction have a non-trivial dependence; however, despite such systematic uncertainties galaxies strongly obscured by dust are still expected to be identifiable via elevated Hα/Hβ line ratios.

Gas-phase oxygen abundances in our sample were derived primarily exploiting the detection of multiple emission lines, where available, in NIRSpec medium-resolution ($R \sim 1000$) grating/filter configurations (G140M/F070LP, G235M/F170LP, G395M/F290LP) taken alongside the PRISM spectroscopic observations (details are discussed in Curti et al.[20]). For targets that were not covered by $R \sim 1000$ observations, the PRISM spectra were considered. More specifically, we required a minimum $3\sigma$ detection on [O III] $\lambda$ 5008 Å, [O II] $\lambda$ 3727, 3730 Å, [Ne III] $\lambda$ 3870 Å, and Hβ to include these lines into the metallicity calculation. On the basis of detected emission lines, we combined the information from the R3, R23, O32, and Ne3O2 line-ratio diagnostics, adopting the calibrations described in Nakajima et al.[63] In the case where only [O III] $\lambda$ 5008 Å and H$\beta$ were detected, and therefore R3 was the only available line ratio, upper limits on [O II] $\lambda$ 3727, 3730 Å and [N II] $\lambda$ 6584 Å were exploited to discriminate between the high- and low-metallicity solutions of the double-branched R3 calibration. The full procedure is described in more detail in Curti et al.[20]. We quote the gas-phase metallicity ($Z_{neb}$) in units of Solar metallicity ($Z_\odot$), assuming $12 + \log_{10}$ (O/H)$_\odot$ = 8.69 as the Solar oxygen abundance[64].

We further explored the rest-frame optical properties of our samples by considering composite spectra around the strong optical emission lines in Extended Data Fig. 3. These stacked spectra were obtained equivalently as will be described in the Spectral stacking section, but with bins of $\Delta\lambda_{emit}$ = 10 Å given the increased spectral resolution of NIRSpec at longer wavelengths[45]. For the purpose of studying the Balmer decrement, we only included galaxies where H$\alpha$ is observable (i.e. we did not consider objects at $z > 7.1$, leaving out one source in the bump sample). We obtained fluxes of the main emission lines (i.e. [O II] $\lambda$ 3727, 3730 Å, [O III] $\lambda$ 4960, 5008 Å, H$\beta$, and H$\alpha$) by fitting Gaussian profiles, shown in Extended Data Fig. 3. Measured line ratios are reported in Extended Data Table 2.

**Stellar population synthesis modelling**

We employed the BAGPIPES code[53] to model the SED simultaneously probed by the NIRSpec PRISM measurements and NIRCam photometry, for which we use a conservative 10% error floor. As underlying stellar models, we used the Binary Population and Spectral Synthesis (BPASS[31]) v2.2.1 stellar population synthesis models including binary stars under the default BPASS IMF, having a slope of −2.35 (for $M > 0.5$ M$_\odot$) and ranging in stellar mass from 1 M$_\odot$ to 300 M$_\odot$. Aiming for a model that is simple yet able to capture older stellar populations, we adopted a constant SFH with a minimum age varying between 0 (i.e. ongoing star formation) and 500 Myr, and a maximum age varying between 1 Myr and the age of the Universe. The total stellar mass formed was varied between 0 and $10^{15}$ M$_\odot$, and the stellar metallicity between 0 and 1.5 Z$_\odot$. Nebular emission was included in a self-consistent manner using a grid of Cloudy[65] models parametrised by the ionisation parameter ($-3 < \log_{10} U < -0.5$). We chose a flexible Charlot & Fall[66] dust attenuation prescription with varying visual extinction ($0 < A_V < 7$ mag) and power-law slope ($0.4 < n < 1.5$), while we fixed the fraction of attenuation arising from stellar birth clouds to 60% (the remaining fraction originating in the diffuse ISM; e.g. ref.[67]). We note the Calzetti et al.[59] dust attenuation curve yields consistent results. A first-order Chebyshev polynomial (described in Carnall et al.[68]) was included to account for aperture and flux-calibration effects in the spectroscopic data. The detailed properties of JADES-GS-z6-0 are reported in Table 1. Moreover, the resulting stellar masses ($M_*$), star formation rates (SFRs) averaged over the last 30 Myr (SFR$_{30}$), and mass-weighted stellar ages ($t_*$) inferred from SED models of the entire sample are presented in Extended Data Fig. 2. Median values of all properties for the galaxy sample with and without evidence for a UV bump are reported in Extended Data Table 2.

**Stellar population age determination**

We further explored whether the apparent absence of a significantly older stellar population (i.e. $t_* > 300$ Myr) could be explained by an "outshining effect" due to a more recent burst of star formation[69]. Indeed, there is evidence that a significant fraction (20% to 25%) of reionisation-era galaxies ($z \gtrsim 6$) host such evolved stellar populations[70,71]. Taking the best-fit parameters in our

BAGPIPES model, we added an instantaneous burst of star formation to the original model with a single (constant) SFH component. Comparing the reduced chi-squared values between the original, single-component model and the new two-component model (accounting for an additional three model parameters, namely stellar mass, metallicity, and age of the burst), we inferred, from a stellar population synthesis modelling point of view, how large a stellar mass can be "disguised" in an evolved stellar population. This is illustrated in Extended Data Fig. 4, showing the age-sensitive 4000 Å (Balmer) break. To avoid systematic uncertainties due to flux calibration and/or slit losses in the spectrum, we restricted the chi-squared analysis to the photometry. We determined the difference in reduced chi-squared values with $\Delta\chi^2_\nu = \chi^2_{\nu, evolved} - \chi^2_{\nu, original}$, where $\chi^2_{\nu, original}$ ($\chi^2_{\nu, evolved}$) is the reduced chi-squared metric of the single-component (two-component) model. From this conservative estimate, we cannot definitively rule out the existence of an additional population of evolved stars; for example, for $\Delta\chi^2_\nu = 4$ (i.e. at $2\sigma$ or 95% confidence) JADES-GS-z6-0 can have up to $5.5 \cdot 10^7$ $M_\odot$ ($9.6 \cdot 10^7$ $M_\odot$), $0.55\times$ ($0.95\times$) its inferred stellar mass, in a 250 Myr (500 Myr) old burst of star formation. This scenario, however, where a galaxy builds up more than half of its stellar mass following an extended period (i.e. more than 250 Myr) with little or no star formation, is physically implausible given the smooth SFH expected for relatively massive galaxies in this early epoch ($M_* \gtrsim 10^8$ $M_\odot$)[72]. Even a more stochastic mode of star formation is not likely to undergo such a lengthy quiescent period, suggesting that the SEDs should reveal detectable signatures of stars with intermediate ages (~100 Myr) if star formation activity can truly be traced back over a time period required for AGB stars to produce significant amounts of dust. Instead, we constrain an additional 100 Myr old component to have at most $0.31\times$ the current stellar mass (~$3 \cdot 10^7$ $M_\odot$; $2\sigma$). This suggests that more than half, if not most, of the stellar mass in JADES-GS-z6-0 was built up in less than 100 Myr. Finally, we note that stacked rest-frame optical spectra (discussed in the Spectroscopic rest-frame optical properties section), when normalised to the continuum at $\lambda_{emit} \sim 3600$ Å, equally do not reveal a strong Balmer break either in the bump sample or in the full sample, further supporting the finding that these galaxies have relatively young stellar populations.

**Ancillary far-infrared observations**

To search for additional signatures of dust obscuration, we considered archival Atacama Large Millimeter/submillimeter Array (ALMA) 1.2 mm and 3 mm continuum imaging taken over GOODS-S. All sources in our sample are contained within the combined 1.2 mm data of the ALMA twenty-six arcmin$^2$ survey of GOODS-S at one millimeter (ASAGAO[73]; ALMA project code 2015.1.00098.S, PI: K. Kohno) survey, which includes the ALMA HUDF[74] (project code 2012.1.00173.S, PI: J. Dunlop) and GOODS-ALMA[75] (project code 2015.1.00543.S, PI: D. Elbaz) surveys and reaches a continuum sensitivity of ~78 μJy ($3\sigma$). A further 15 sources, including three sources in the bump sample (JADES-GS+53.17022-27.77739, JADES-GS+53.16743-27.77548, and JADES-GS+53.16660-27.77240), are covered by the ALMA SPECtroscopic Survey (ASPECS[76,77]; project code 2013.1.00146.S, PI: F. Walter), reaching a $3\sigma$ continuum sensitivity of ~38 μJy at 1.2 mm and ~11.4 μJy at 3 mm. None of the 49 sources in our sample, however, show a significant detection ($3.5\sigma$) in either dataset. A stacking procedure similarly does not yield any detectable continuum emission, neither for the sources in the bump sample nor for the full sample, indicating that the non-detections can be explained by the relatively low sensitivity of the ALMA mosaics. Indeed, we have verified that for a typical

SFR of a few solar masses per year (as inferred for JADES-GS-z6-0), even a conservatively high fraction (50%) of dust-obscured star formation results in an infrared luminosity that requires several tens of hours to secure a confident detection ($L_{IR} \sim 10^{10}\ L_\odot$, translating into a continuum flux density of $F_\nu \sim 5\ \mu Jy$ in band 6).

## Bump parametrisation and fitting procedure

Given an observed flux density profile $F_\lambda$, we parametrised the UV bump profile by defining the excess attenuation as in Shivaei et al.[13], $A_{\lambda,\ bump} = -2.5 \log_{10}(F_\lambda/F_{\lambda,\ cont})$. For the individual spectrum of JADES-GS-z6-0, we took the power-law fit with UV slope $\beta_{UV}$ measured outside the bump region as the attenuated spectrum without a bump, $F_{\lambda,\ cont}$. When considering the excess attenuation in the individual spectrum of JADES-GS-z6-0, we again used the running median and corresponding uncertainty (described in the Sample selection section), which was additionally used to compute the significance of the negative flux excess of the spectrum with respect to the power-law fit alone (panel c of Fig. 1). We note the formal uncertainty of each spectral pixel is scaled upwards to include the effects of covariance between adjacent pixels; we have verified a similarly high significance is found when bootstrapping a spectrum first rebinned to match the spectral resolution element (thereby largely negating the effects of correlated noise). Using the MultiNest[56] nested sampling algorithm, we fitted the excess attenuation $A_{\lambda,\ bump}$ with a Drude profile[78], which has been shown to appropriately describe the spectral shape of the bump[13,18,79]. Centred on rest-frame wavelength $\lambda_{max}$, it is parametrised as

$$A_{\lambda,\ bump} = A_{\lambda,\ max} \frac{\gamma^2/\lambda^2}{\left(1/\lambda^2 - 1/\lambda_{max}^2\right)^2 + \gamma^2/\lambda^2}$$

where the full width at half maximum (FWHM) is given by FWHM $= \gamma \lambda_{max}^2$. We fixed $\gamma = 250$ Å$/(2175$ Å$)^2$ which, if $\lambda_{max} = 2175$ Å, corresponds to FWHM $= 250$ Å in agreement with what has been found for $z \sim 2$ star-forming galaxies[13,18]. Again motivated by the spectral windows defined by Calzetti et al.[59], we performed the fitting procedure in a region of 1950 Å $\leq \lambda_{emit} \leq 2580$ Å (reflecting the $\gamma_3$ and $\gamma_4$ regions discussed in the Sample selection section), which excludes the C III doublet. As a prior for the bump amplitude $A_{\lambda,\ max}$, we conservatively chose a gamma distribution with shape parameter $a = 1$ and scale $\theta = 0.2$, which favours the lowest amplitudes (noting a flat prior yields comparable results). For the central wavelength, we adopted a flat prior in the range 2100 Å $< \lambda_{max} < 2300$ Å.

## Spectral stacking

To perform a spectral stacking analysis, we shifted each spectrum to rest-frame wavelengths $\lambda_{emit}$ and normalised it to the value of the power-law fit at $\lambda_{emit} = 2175$ Å. The individual continuum spectra and corresponding uncertainties are rebinned to bins of $\Delta\lambda_{emit} = 20$ Å using SpectRes[80]. Stacked continuum profiles were created by weighting each binned data point by its inverse variance, though we note we obtain similar results with an unweighted average. The stacked continuum profile $F_\lambda$ of the ten galaxies with evidence for a UV bump is converted to an excess attenuation as described in the previous section, where for the "bumpless" profile ($F_{\lambda,\ cont}$), we refitted a power-law continuum to the stacked continuum profile of the ten galaxies, noting the difference in slope (measured to be $\beta \sim -1.95$) compared to the stacked spectrum of the full sample of 49 galaxies (with $\beta \sim -2.12$, see Extended Data Table 2). To ensure good agreement with the observed continuum outside of the region used in the bump fitting procedure, this power-law was determined from the Calzetti et al.[59] windows bluewards of $\lambda_{emit} = 1850$ Å (explicitly excluding the C III doublet region), whereas at wavelengths beyond the bump region, we consider the windows 2500 Å $< \lambda_{emit} <$ 2600 Å and 2850 Å $< \lambda_{emit} <$ 3000 Å (avoiding potential Mg II doublet emission at $\lambda_{emit} \sim 2800$ Å). Fitting a Drude profile[78] yields an amplitude of $0.10^{+0.01}_{-0.01}$ mag and a central wavelength $\lambda_{max} = 2236^{+21}_{-20}$ Å. We note the amplitude remains effectively unchanged if we instead fix the central wavelength to $\lambda_{max} = 2175$ Å.

## Robustness of the UV bump detections

### JADES-GS-z6-0

To test the robustness of the UV bump identification in JADES-GS-z6-0, we extracted one-dimensional spectra from the three separate observing visits to show that the feature around 2175 Å is not dominated by a single observation. This is illustrated in Extended Data Fig. 5, which shows measurements from the individual visits normalised to its power-law fit. We furthermore tested our extraction of the one-dimensional spectra using different apertures on the reduced two-dimensional spectra (see the Data and parent sample section). This slightly lowers the average continuum flux level and SNR, but we find no significant changes to the rest-frame UV spectrum. We also compared NIRCam apodised photometry (the total background-subtracted NIRCam flux passing through the NIRSpec MSA slit) to synthetic photometry obtained from convolving the PRISM spectra with NIRCam filters. We verified that for most sources the two fluxes are offset by a constant factor with offsets smoothly varying as a function of wavelength. Finally, we note the attenuation feature is a highly localised region in the low-resolution PRISM spectra (a rest-frame width of 250 Å is sampled by ~6 independent spectral resolution elements at a resolution of $R(\lambda_{obs} \sim 1.7~\mu m) \sim 50$) such that its magnitude will not be significantly affected by the absolute flux calibration. At the same time, this wavelength range is probed by more than 10 native detector pixels, indicating the chances that this feature is produced by correlated detector noise or other artefacts are minimal.

### Stacked spectra

In this section, we discuss the robustness of the identification of the bump feature in our stacked spectra. Firstly, randomly splitting our bump sample into two subsamples, we have confirmed the bump signature remains present in both, implying the stacked spectrum is not dominated by a single source. Further, we have verified that performing an analogous stacking procedure at a different wavelength (2475 Å), for a subset of sources selected based on the continuum shape around 2475 Å in an equivalent manner as the $\gamma_{34}$ selection described in the Sample selection section, does not produce a significant broad absorption feature as in Fig. 2. Instead, this results in a narrow negative excess with positive excess on the edge of our fitting window, hence yielding a substantially lower amplitude when fitted with a Drude profile.

We now turn to explore various properties of the different samples measured by NIRSpec and NIRCam to test whether the bump signature could purely be due to random noise fluctuations, in which case the ten selected galaxies are expected to simply be a random subset of the parent sample. As seen in Extended Data Fig. 2, we find a significant correlation (i.e. $p < 0.05$ for the null hypothesis that the data is uncorrelated) between on the one hand the UV slope inflection around 2175 Å, $\gamma_{34}$, and on the other hand absolute UV magnitude $M_{UV}$. Our selected bump sample is measured to be intrinsically fainter in the rest-frame UV (higher $M_{UV}$, independent of the SED modelling). This may be indicative of the absence of young stellar populations, in line with the theoretically predicted trend of decreasing bump strength with increasing star formation activity, and hence the intensity and hardness of UV radiation field[81]. Moreover, several of the median properties hint at systematically different physical conditions in galaxies part of the bump sample: in particular, these objects exhibit a significantly enhanced Hα/Hβ ratio, indicating that on average the nebular emission in these galaxies experiences a higher degree of dust obscuration, with nebular extinction values comparable to those of $z \sim 2$ galaxies with a UV bump[18]. Moreover, their slightly elevated gas-phase oxygen abundances indicate that they are more metal enriched (see Extended Data Table 2). Interestingly, however, the stellar masses of the bump sample are significantly lower than their $z \sim 2$ counterparts, as pointed out in Fig. 3. We note that other factors such as geometry could play an important role in determining the strength of the UV bump, though larger samples are needed to confirm these trends.

To avoid potential biases by contaminants in our $\gamma_{34}$-selected sample in the correlations based on individual galaxy properties, we further turn to explore the stacked spectra. For instance, we note the bump and non-bump samples appear to be characterised by a comparable median UV slope as measured in the individual spectra, which is confirmed by the agreement of UV slopes in *unweighted* stacked spectra. However, the *weighted* stacked spectrum shown in Fig. 2 reveals that the bump sample has a significantly redder UV continuum (as discussed in the Spectral stacking section). From the stacked spectra around the strong optical emission lines in Extended Data Fig. 3, we again find the Hα/Hβ ratio in the bump sample to be significantly higher, translating into a nebular extinction $E(B-V)_{neb}$ a factor of ~2 higher than in the stacked spectrum of the full sample. This indicates the bump sample preferably contains dustier galaxies, strongly favouring the interpretation that the observed excess attenuation around 2175 Å is due to dust absorption. Moreover, we find evidence for a mildly higher metallicity in the bump sample through an enhanced line ratio of [O III] $\lambda$ 5008 Å to Hβ. Though this line ratio follows a double-branched metallicity solution (e.g. ref.[20]), the low-metallicity solution that monotonically increases with [O III]/Hβ should be appropriate for the current sample of galaxies, given the [O III]/[O II] line ratio of approximately 10 (both in the full sample and the subset of sources selected in the bump sample). We note such differences in the Hα/Hβ and [O III]/Hβ line ratios are absent in the control sample (selected based on the continuum shape around 2475 Å) discussed above.

Finally, we have verified that we are able to perform a blind selection from the parent sample of sources with the highest Balmer decrements and reddest UV slopes which results in a tentative detection of the UV bump. Specifically, requiring a Balmer decrement of Hα/Hβ ≳ 4 and UV slope of $\beta_{UV} \gtrsim -2.2$ yields a sample of four sources all contained within the bump sample (i.e. JADES-GS+53.16871-27.81516, JADES-GS+53.13284-27.80186, JADES-GS+53.17022-27.77739, and JADES-GS+53.16743-27.77548; see Extended Data Table 1) but notably excludes JADES-GS-z6-0. Without any pre-selection on the continuum shape around 2175 Å, the stacked spectrum of these four galaxies produces a tentative (~4$\sigma$) bump feature.

## Bump amplitude comparison with literature results

As discussed in Shivaei et al.[13], the adopted parametrisation of bump amplitude in the excess attenuation (i.e. $A_{\lambda,\,max}$; see the Bump parametrisation and fitting section) includes the extinction in the absence of the bump, $E(B-V)$, to avoid propagating the large uncertainties of this parameter that stem from not well-constrained assumptions on the attenuation curves of high-redshift galaxies. We note a direct measurement of the Balmer recombination line ratios can in principle constrain the nebular extinction[82], but its relation with stellar extinction carries uncertainty in addition to suffering from wavelength-dependent slit-loss effects (also discussed in the Spectroscopic rest-frame optical properties section). In Fig. 3, we directly compare these excess attenuation strengths, taking into account the underlying extinction $E(B-V)$ for bump strengths measured in MW, LMC, and SMC extinction curves. In terms of the commonly used Fitzpatrick & Massa[40,78,83] parametrisation, $A_{\lambda,\,max} = c_3/\gamma^2\,E(B-V)$. We retrieve $E(B-V)$ from the measured total-to-selective extinction (i.e. $R_V = A_V / E(B-V)$) in each extinction curve, assuming a range of $0.1$ mag $< A_V < 0.5$ mag. Data points from Noll et al.[18] and Heintz et al.[39] (and references therein) are similarly converted to a consistent bump amplitude $A_{\lambda,\,max}$ using their measured values of $E(B-V)$. Measurements from Noll et al.[18] represent the stacked spectra of three subsamples that were selected based on their UV slope $\beta_{UV}$ and bump strength parametrised by the $\gamma_{34}$ parameter (see the Sample selection section): the upwards pointing triangle in Fig. 3 has $\beta_{UV} < -1.5$ and $\gamma_{34} > -2$, the diamond has $\beta_{UV} > -1.5$ and $\gamma_{34} > -2$, and the downward pointing triangle has $\gamma_{34} < -2$. We note that the Heintz et al.[39] measurements of gamma-ray burst absorbers are effectively along a line of sight through the galaxies, while the Shivaei et al.[13] and Noll et al.[18] measurements, similar to the measurements in this work, are based on the total integrated light of galaxies. The distribution of dust with respect to the stars within galaxies affects the latter, integrated observations of the UV bump[24,84].

## References (continued)

## Data availability

The data that support the findings of this study are publicly available[48,51] at https://archive.stsci.edu/hlsp/jades.

## Code availability

The AstroPy[85,86] software suite is publicly available, as is BAGPIPES[53], MultiNest[56], PyMultiNest[57], pPXF[60], and SpectRes[80]. BEAGLE[51] is available via a Docker image upon request at http://www.iap.fr/beagle/.

## Acknowledgements

This work is based on observations made with the NASA/ESA/CSA *James Webb Space Telescope*. The data were obtained from the Mikulski Archive for Space Telescopes at the Space Telescope Science Institute, which is operated by the Association of Universities for Research in Astronomy, Inc., under NASA contract NAS 5-03127 for *JWST*. These observations are associated with programmes 1180, 1210, 1895, and 1963. The authors acknowledge the FRESCO team led by PI Pascal Oesch for developing their observing program with a zero-exclusive-access period. JW gratefully acknowledges support from the Fondation MERAC. JW, RM, MC, FDE, TJL, WMB, LS, and JS acknowledge support by the Science and Technology Facilities Council (STFC), by the European Research Council (ERC) through Advanced Grant 695671, "QUENCH", and by the UK Research and Innovation (UKRI) Frontier

Research grant RISEandFALL. RS acknowledges support from an STFC Ernest Rutherford Fellowship (ST/S004831/1). RM also acknowledges funding from a research professorship from the Royal Society. SCa acknowledges support by the European Union's HE ERC Starting Grant No. 101040227 – WINGS. ECL acknowledges support of an STFC Webb Fellowship (ST/W001438/1). SAr, BRDP, and MP acknowledge support from the research project PID2021-127718NB-I00 of the Spanish Ministry of Science and Innovation/State Agency of Research (MICIN/AEI). AJB, AJC, JC, and AS acknowledge funding from the "FirstGalaxies" Advanced Grant from the European Research Council (ERC) under the European Union's Horizon 2020 research and innovation programme (Grant agreement No. 789056). SAl acknowledges support from the *JWST* Mid-Infrared Instrument (MIRI) Science Team Lead, grant 80NSSC18K0555, from NASA Goddard Space Flight Center to the University of Arizona. EE, DJE, BDJ, MR, BER, FS, and CNAW acknowledge the *JWST*/NIRCam contract to the University of Arizona, NAS5-02015. DJE is supported as a Simons Investigator. MP acknowledges support from the Programa Atracción de Talento de la Comunidad de Madrid via grant 2018-T2/TIC-11715. CCW is supported by NOIRLab, which is managed by the Association of Universities for Research in Astronomy (AURA) under a cooperative agreement with the National Science Foundation. This research is supported in part by the Australian Research Council Centre of Excellence for All Sky Astrophysics in 3 Dimensions (ASTRO 3D), through project number CE170100013. This study made use of the Prospero high performance computing facility at Liverpool John Moores University. The authors acknowledge use of the lux supercomputer at UC Santa Cruz, funded by NSF MRI grant AST 1828315.

**Author contributions**

JW, IS, RS, and RM led the writing of this paper. MR, EE, FS, KNH, and CCW contributed to the design, construction, and commissioning of NIRCam. AJB, CNAW, CW, DJE, MR, PF, RM, SAl, and SAr contributed to the design of the JADES survey. BER, ST, BDJ, CNAW, DJE, IS, RE, SAl, and ZJ contributed to the JADES imaging data reduction. BER contributed to the JADES imaging data visualisation. BDJ, ST, RE, EN, and WMB contributed to the modelling of galaxy photometry. KNH, JL and RE contributed the photometric redshift determination and target selection. BER, CNAW, CCW, KNH, and MR contributed to the JADES pre-flight imaging data challenges. SCa, MC, JW, PF, GG, SAr, MP, and BRdP contributed to the NIRSpec data reduction and to the development of the NIRSpec pipeline. SAr contributed to the design and optimisation of the MSA configurations. AJC, AJB, CNAW, ECL, and KB contributed to the selection, prioritisation and visual inspection of the targets. SCh, JC, ECL, RM, JW, RS, FDE, TJL, MC, AdG, AS, and LS contributed to analysis of the spectroscopic data, including redshift determination and spectral modelling. PF, TR, GG, NK, and BRdP contributed to the design, construction and commissioning of NIRSpec. FDE, TJL, MC, BRdP, RM, SA, and JS contributed to the development of the tools for the spectroscopic data analysis, visualisation and fitting. CW contributed to the design of the spectroscopic observations and MSA configurations. BER, CW, DJE, MR, and RM serve on the JADES Steering Committee.

**Competing interests**

The authors declare no competing interests.

## Additional Information

| Source name | $z_{\text{spec}}$ | $m_{\text{F444W}}$ (mag) | $M_{\text{UV}}$ (mag) | $\beta_{\text{UV}}$ | H$\alpha$/H$\beta$ |
|---|---|---|---|---|---|
| JADES-GS+53.15138-27.81917† | $6.7065^{+0.0004}_{-0.0003}$ | $28.23 \pm 0.06$ | $-18.34 \pm 0.12$ | $-2.13^{+0.18}_{-0.18}$ | $3.67 \pm 0.24$ |
| JADES-GS+53.16871-27.81516‡ | $4.0223^{+0.0002}_{-0.0002}$ | $26.17 \pm 0.01$ | $-19.58 \pm 0.06$ | $-1.85^{+0.06}_{-0.06}$ | $4.76 \pm 0.07$ |
| JADES-GS+53.13059-27.80771 | $5.6166^{+0.0009}_{-0.0007}$ | $27.88 \pm 0.04$ | $-18.74 \pm 0.07$ | $-2.02^{+0.20}_{-0.21}$ | $2.48 \pm 0.41$ |
| JADES-GS+53.13284-27.80186‡ | $4.6617^{+0.0001}_{-0.0001}$ | $26.71 \pm 0.01$ | $-18.85 \pm 0.02$ | $-1.56^{+0.09}_{-0.08}$ | $4.13 \pm 0.05$ |
| JADES-GS+53.17022-27.77739‡ | $4.7094^{+0.0005}_{-0.0004}$ | $27.21 \pm 0.03$ | $-18.99 \pm 0.04$ | $-2.14^{+0.14}_{-0.14}$ | $5.27 \pm 0.76$ |
| JADES-GS+53.12689-27.77689 | $4.8878^{+0.0002}_{-0.0002}$ | $28.38 \pm 0.04$ | $-18.71 \pm 0.04$ | $-2.34^{+0.10}_{-0.09}$ | $3.46 \pm 0.33$ |
| JADES-GS+53.16743-27.77548‡ | $4.1426^{+0.0002}_{-0.0002}$ | $27.55 \pm 0.02$ | $-17.91 \pm 0.08$ | $-2.05^{+0.16}_{-0.16}$ | $4.35 \pm 0.51$ |
| JADES-GS+53.16660-27.77240 | $6.3296^{+0.0002}_{-0.0002}$ | $27.98 \pm 0.05$ | $-19.17 \pm 0.06$ | $-2.35^{+0.13}_{-0.13}$ | $2.99 \pm 0.11$ |
| JADES-GS+53.13423-27.76891 | $7.0493^{+0.0003}_{-0.0003}$ | $28.12 \pm 0.02$ | $-18.86 \pm 0.03$ | $-2.72^{+0.17}_{-0.18}$ | $2.11 \pm 0.14$** |
| JADES-GS+53.11833-27.76901 | $7.2043^{+0.0002}_{-0.0002}$ | $27.77 \pm 0.08$ | …* | $-2.53^{+0.15}_{-0.15}$ | …** |

**Extended Data Table 1 | Properties of the galaxies in the bump sample.**
Error bars on measurements represent 1σ uncertainty. Rows: (1) Source name (including the J2000 RA and Dec in deg), (2) Spectroscopic redshift ($z_{\text{spec}}$), (3) Apparent AB magnitude in the NIRCam F444W filter ($m_{\text{F444W}}$), (4) Absolute AB magnitude in the UV ($M_{\text{UV}}$), (5) UV spectral slope ($\beta_{\text{UV}}$), (6) Balmer decrement (H$\alpha$/H$\beta$).
† JADES-GS-z6-0.
‡ Contained in the blind selection discussed in the Robustness of the UV bump detections section.
∗ This source falls outside the main NIRCam footprint.
∗∗ At this redshift, H$\alpha$ (partially) shifts outside of the NIRSpec coverage.

|  | Bump sample | Non-bump sample | Full sample |
| --- | --- | --- | --- |
| *Individual properties* | | | |
| $z_{\rm spec}$ | $5.25 \pm 0.68$ | $5.77 \pm 0.27$ | $5.62 \pm 0.29$ |
| $M_{\rm UV}$ (mag) | $-18.85 \pm 0.14$ | $-19.17 \pm 0.13$ | $-19.14 \pm 0.13$ |
| $\beta_{\rm UV}$ | $-2.14 \pm 0.12$ | $-2.08 \pm 0.07$ | $-2.12 \pm 0.06$ |
| $\gamma_{34}$ | $-1.86 \pm 0.21$ | $0.62 \pm 0.26$ | $-0.03 \pm 0.34$ |
| $Z_{\rm neb}$ ($Z_\odot$) | $0.17 \pm 0.05$ | $0.10 \pm 0.02$ | $0.11 \pm 0.02$ |
| $E(B-V)_{\rm neb}$ (mag) | $0.25 \pm 0.13$ | $0.06 \pm 0.03$ | $0.06 \pm 0.03$ |
| $M_*$ ($10^8$ $M_\odot$) | $1.8 \pm 0.8$ | $2.3 \pm 0.7$ | $2.3 \pm 0.4$ |
| $Z_*$ ($Z_\odot$) | $0.06 \pm 0.03$ | $0.17 \pm 0.05$ | $0.13 \pm 0.04$ |
| $\rm SFR_{30}$ ($M_\odot$ yr$^{-1}$) | $1.9 \pm 1.6$ | $4.1 \pm 1.0$ | $3.0 \pm 0.8$ |
| $t_*$ (Myr) | $22 \pm 24$ | $29 \pm 5$ | $27 \pm 4$ |
| *Stack properties* | | | |
| $N_{\rm sample}$ | 10 | 39 | 49 |
| $\beta$ | $-1.95^{+0.02}_{-0.02}$ | ... | $-2.18^{+0.01}_{-0.01}$ |
| H$\alpha$/H$\beta$ | $3.96^{+0.05}_{-0.05}$ | ... | $3.33^{+0.02}_{-0.02}$ |
| $E(B-V)_{\rm neb}$ (mag) | $0.33^{+0.01}_{-0.01}$ | ... | $0.15^{+0.01}_{-0.01}$ |
| [O III] $\lambda$ 5008 Å/H$\beta$ | $6.57^{+0.08}_{-0.08}$ | ... | $5.15^{+0.03}_{-0.03}$ |
| [O III] $\lambda$ 5008 Å/[O II] | $11.3^{+0.3}_{-0.3}$ | ... | $11.1^{+0.2}_{-0.2}$ |

**Extended Data Table 2 | Properties of the samples of galaxies with and/or without indication of a UV bump.**

Properties of the samples are presented as median values; error bars represent a 1σ uncertainty (obtained with bootstrapping for the median properties). Rows of individual properties: (1) Spectroscopic redshift ($z_{\rm spec}$), (2) Absolute AB magnitude in the UV ($M_{\rm UV}$), (3) UV spectral slope ($\beta_{\rm UV}$), (4) Spectral slope change around $\lambda_{\rm emit} = 2175$ Å ($\gamma_{34}$), (5) Gas-phase metallicity ($Z_{\rm neb}$; from rest-frame optical emission lines) in units of Solar metallicity, (6) Nebular extinction ($E(B-V)_{\rm neb}$; from the Balmer decrement) in magnitudes, (7) Stellar mass ($M_*$) in $10^8$ Solar masses, (8) Stellar metallicity ($Z_*$; from SED modelling) in units of Solar metallicity, (9) Star formation rate in Solar masses per year averaged on a timescale of 30 Myr ($\rm SFR_{30}$), (10) Mass-weighted stellar age ($t_*$) in Myr. Rows of properties measured in stacked spectra: (1) Number of galaxies contained within the sample ($N_{\rm sample}$), (2) Power-law slope ($\beta$), (3) Balmer decrement H$\alpha$/H$\beta$, (4) Nebular extinction ($E(B-V)_{\rm neb}$) in magnitudes, (5) [O III] $\lambda$ 5008 Å/H$\beta$ line ratio, (6) [O III] $\lambda$ 5008 Å/[O II] line ratio.

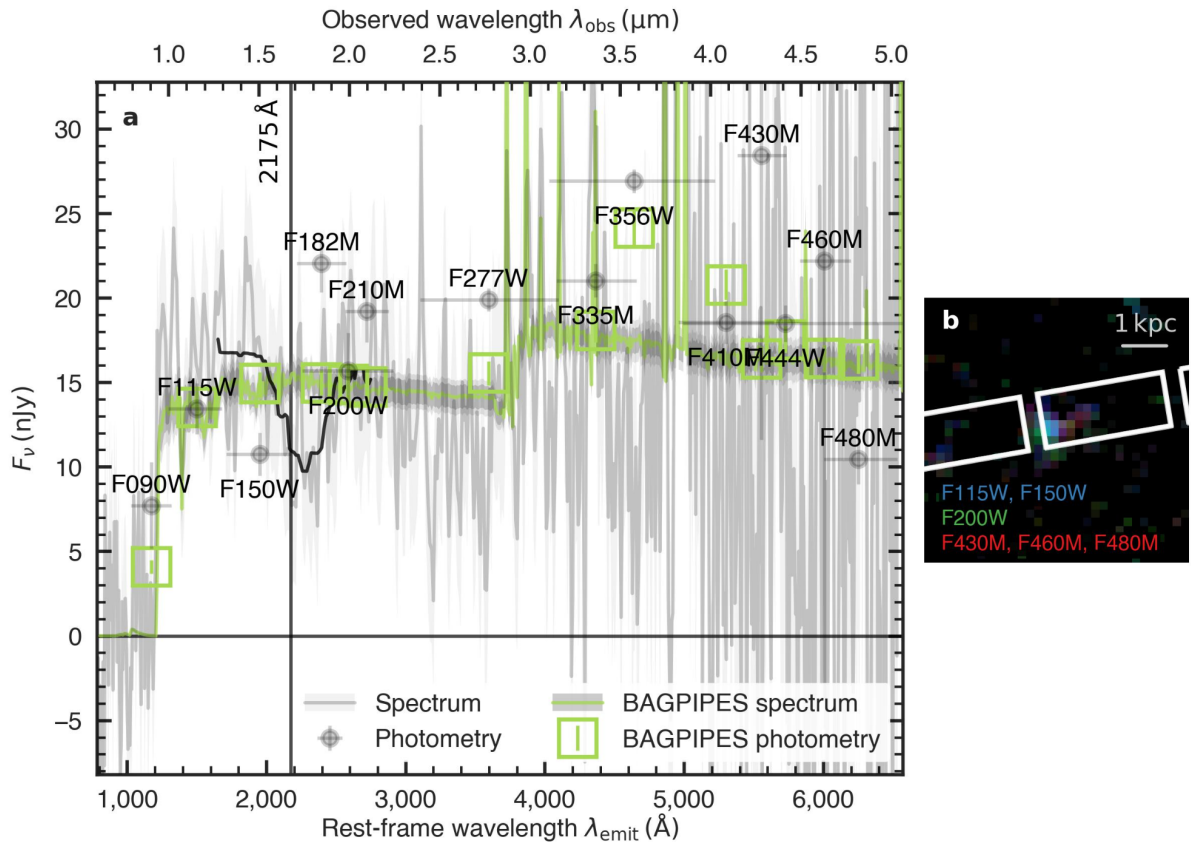

**Extended Data Fig. 1 | SED modelling and false-colour image of JADES-GS-z6-0. a**, Spectrum observed with NIRSpec (grey solid line and light-grey shading as 1σ uncertainty), overlaid with NIRCam photometry (grey points; error bars show the filter widths along the wavelength direction and 1σ uncertainty along the y-axis) of JADES-GS-z6-0. A calibration correction is applied to the spectrum to match the photometry (see the Stellar population synthesis modelling section). The running median around 2175 Å is shown as a solid black line. The best-fit bagpipes SED model (Stellar population synthesis modelling section) is shown as a light green solid line (predicted spectrum; darker and lighter shading represents the 1σ and 2σ uncertainty, respectively) and points (predicted photometry; error bars show 1σ uncertainty). **b**, Colour-composite 1″ × 1″ image constructed from inverse-variance weighted combinations of NIRCam filters, with the F115W and F150W filters as the blue channel, the F200W filter as green, and three medium-band filters not contaminated by strong emission lines (F430M, F460M, and F480M) as red. The position of the NIRSpec MSA shutters in the central nodding position (shown for all three separate observing visits, which however are nearly identical) are overlaid in white. A scale of 1 kpc is indicated in the bottom right.

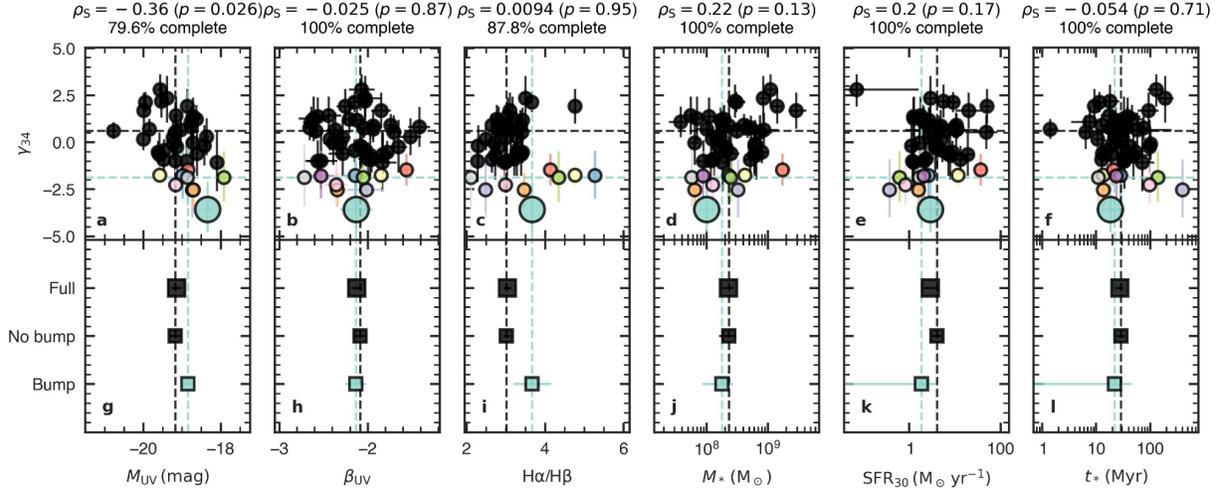

**Extended Data Fig. 2 | Sample characteristics.** Top row (panels **a** to **f**): spectral slope change around $\lambda_{emit} = 2175$ Å ($\gamma_{34}$) as a function of the physical properties of the galaxy sample. JADES-GS-z6-0 is highlighted as an enlarged point. Error bars represent $1\sigma$ uncertainty. The median values of galaxies belonging to the sample with (coloured points) and without (black points) a UV bump signature (see the Sample selection section) are indicated respectively with coloured and black dashed lines. The Spearman's rank correlation coefficient, $\rho_S$, is reported at the top of each panel along with its $p$-value and the completeness (i.e. for which percentage of the sample the metric was measured). Bottom row (panels **g** to **l**): median of the full sample (large black square), the sample with (coloured square) and without (small black square) bump signatures. The error bars show uncertainties on the median obtained with bootstrapping. Quantities shown are the absolute UV magnitude ($M_{UV}$; panels **a** and **g**), UV spectral slope ($\beta_{UV}$; **b** and **h**), Balmer decrement H$\alpha$/H$\beta$ (**c** and **i**), stellar mass ($M_*$; **d** and **j**), star formation rate averaged over the last 30 Myr (SFR$_{30}$; **e** and **k**), and mass-weighted stellar age ($t_*$; **f** and **l**).

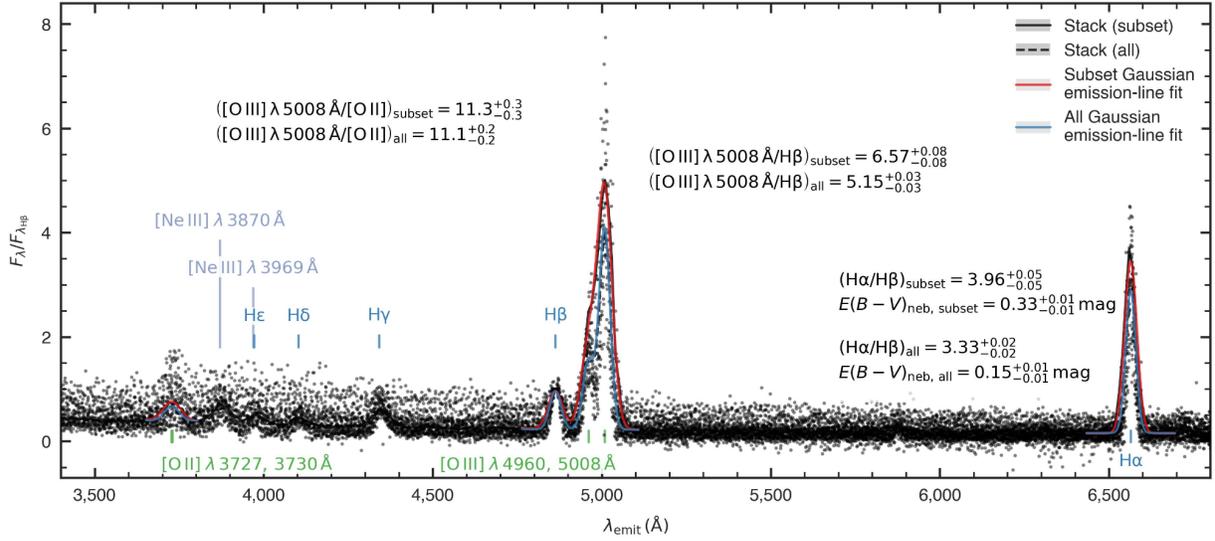

**Extended Data Fig. 3 | Normalised and stacked rest-frame optical spectra of z > 4 JADES galaxies observed by *JWST*/NIRSpec.** Similar to Fig. 2, spectra of all galaxies (small black dots) are shifted to the rest frame but normalised to the flux density at the rest-frame wavelength of Hβ, $\lambda_{H\beta} = 4862.7$ Å. The dashed black line (shading represents 1σ uncertainty) shows a stacked spectrum obtained by combining all 49 objects in wavelength bins of $\Delta\lambda_{emit} = 10$ Å. The main rest-frame optical emission lines are labelled. The stacked spectrum of ten galaxies selected to have a bump signature (solid black line, shading as 1σ uncertainty) exhibits stronger [O III] and Hα emission relative to the full sample, while having a consistent [O III]/[O II] line ratio. This is quantified by the integrated flux ratios of the fitted Gaussian line profiles (respectively shown as red and blue solid lines with shading as 1σ uncertainty), indicating differences in oxygen abundance as well as dust obscuration.

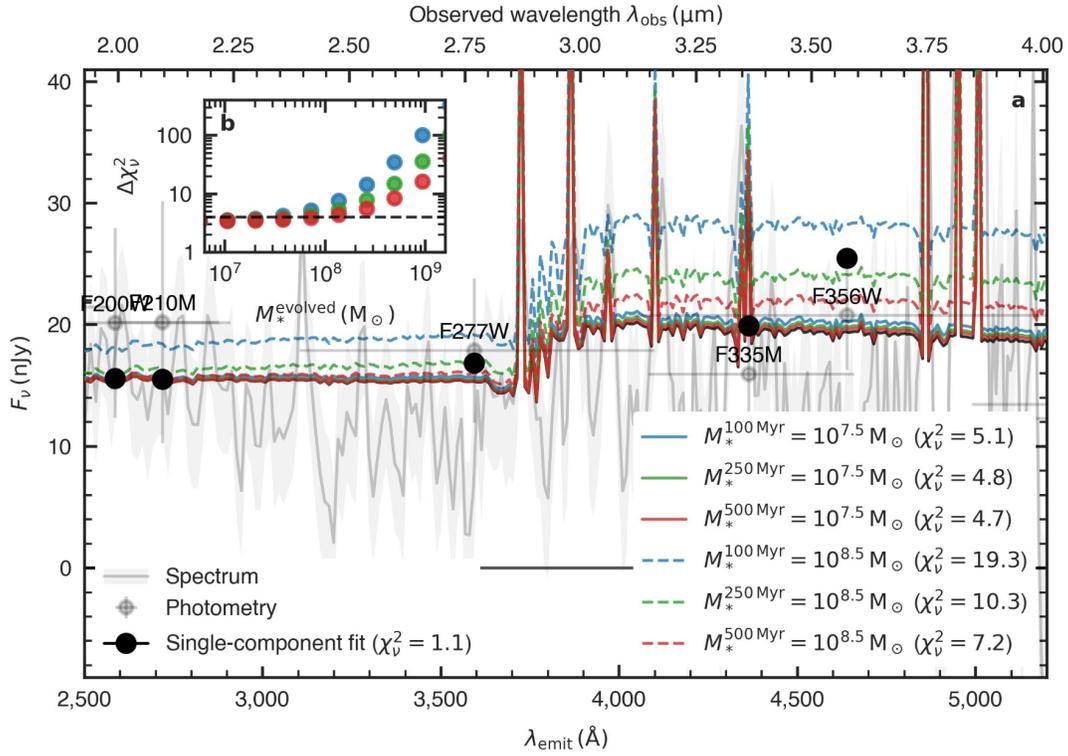

**Extended Data Fig. 4 | Exploration of signatures from significantly evolved stellar populations. a**, The age-sensitive spectral region around the 4000 Å (Balmer) break of JADES-GS-z6-0. Similar to Extended Data Fig. 1, the (photometry-corrected) NIRSpec spectrum is shown as the grey solid line (light-grey shading as $1\sigma$ uncertainty), NIRCam photometry as grey points (error bars show the filter widths along the wavelength direction and $1\sigma$ uncertainty along the y-axis). The BAGPIPES SED model with best-fit parameters (see Stellar population synthesis modelling) is shown as a black solid line (spectrum) and points (photometry). Two-component models are shown by the blue (age of 100 Myr), green (250 Myr), and red (500 Myr) solid ($10^{7.5}$ M$_\odot$) and dashed ($10^{8.5}$ M$_\odot$) lines. **b**, The difference in reduced chi-squared values, $\Delta\chi^2_\nu$, is shown as a function of the stellar mass of the evolved population, $M_{\rm evolved}$. Dashed horizontal lines indicate the value above which the new model is in $2\sigma$ tension with respect to the single-component model ($\Delta\chi^2_\nu = 4$).

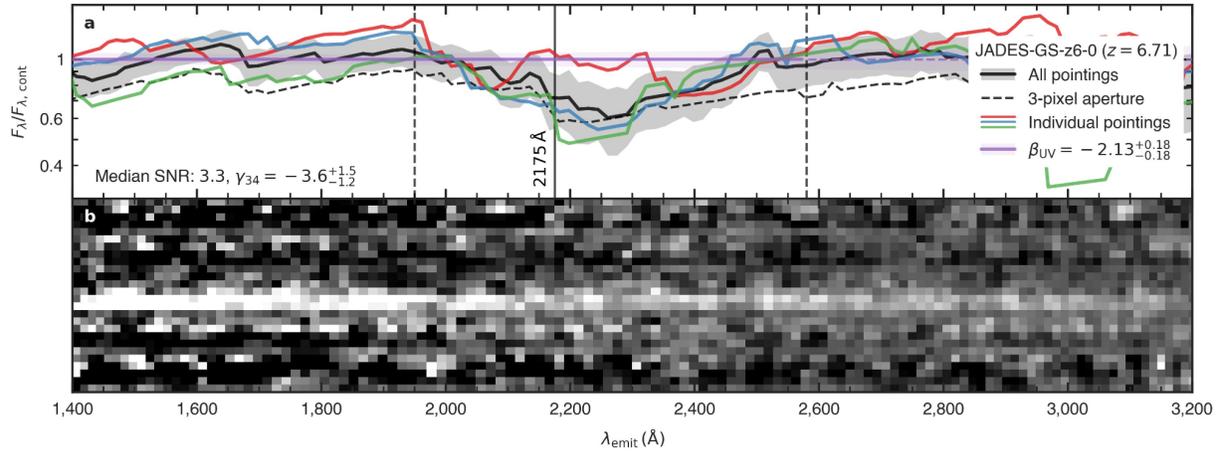

**Extended Data Fig. 5 | Rest-frame UV continuum of JADES-GS-z6-0. a**, The one-dimensional spectrum of JADES-GS-z6-0 (smoothed with a 15-pixel median filter as in Fig. 1) is normalised to the predicted continuum level in the absence of a UV bump, modelled as a power law with index $\beta_{UV}$ (purple line; shading as $1\sigma$ uncertainty). The median SNR and $\gamma_{34}$ are reported in the bottom-left corner. Coloured lines show data from individual observing visits, while the solid black line and grey shading represent the combined spectra and their $1\sigma$ uncertainty, respectively. A dashed black line indicates the spectrum from a 3-pixel aperture extraction. **b**, Two-dimensional spectrum of JADES-GS-z6-0 (not scaled to the predicted continuum level).